\documentclass[reprint,superscriptaddress,showpacs,preprintnumbers,amsmath,amssymb,showkeys]{revtex4-1}
\usepackage{graphicx}
\usepackage{amsmath}
\usepackage[final]{hyperref}
\usepackage{amssymb}
\usepackage{textcomp}

\begin{document}

\preprint{published in: J. Phys. Chem. A, 2009, 113 (37), pp 9952-9957. \href{DOI: 10.1021/jp905039f}{DOI: 10.1021/jp905039f}}
\title[Marksteiner et al.]{UV and VUV ionization of organic molecules, clusters and complexes}

\author{Markus Marksteiner}
\author{Philipp Haslinger}
\author{Michele Sclafani}
\affiliation{Faculty of Physics, University of Vienna,
Boltzmanngasse 5, A-1090 Vienna, Austria}
\author{Hendrik Ulbricht}
\affiliation{School of Physics and Astronomy, University of Southampton, Highfield, Southampton, SO171BJ, United Kingdom}
\author{Markus Arndt}
\email{markus.arndt@univie.ac.at}
\homepage{http://www.quantumnano.at}
\affiliation{Faculty of Physics, University of Vienna,
Boltzmanngasse 5, A-1090 Vienna, Austria}

\date{\today}
\begin{abstract}
The generation of organic particle beams is studied in combination
with photoionization using uv radiation at 266\,nm and vuv light at
157\,nm. Single-photon ionization with pulsed vuv light turns out to
be sensitive enough to detect various large neutral biomolecular
complexes ranging from metal-amino acid complexes to nucleotide
clusters and aggregates of polypeptides. Different biomolecular
clusters are shown  to exhibit rather specific binding
characteristics with regard to the various metals that are
co-desorbed in the source. We also find that the ion signal of
gramicidin can be increased by a factor of fifteen when the photon
energy is increased from 4.66\,eV to 7.9\,eV.
\end{abstract}

\maketitle

\section{Introduction}
Photoionization has become a  widely used technique for the
detection and analysis of neutral molecular beams. For many
molecules of biological relevance the ionization potential lies
between 7--12\,eV. This energy may for instance be provided  in a
multi-photon
process\,\cite{Dey1991a,Koester1992a,Aicher1995a,Weinkauf2002a},
which is often implemented using near-resonant two-photon absorption
at $\lambda=250-300$\,nm. For various selected molecules
single-photon excitation may also be achieved using vacuum
ultra-violet (vuv) laser radiation as generated by excimer lasers
\cite{Arps1989a,Koester1992a,Butcher1999a,Muehlberger2005a}, by
synchrotron radiation\,\cite{Vries1992a} or by high harmonic
generation of strong infrared lasers\,\cite{Shi2002a}.

Photoionization of neutral organic molecules is not only of general
interest for analytic mass spectrometry, but also for matter wave
experiments aiming at either an improved understanding of quantum
coherence and decoherence
\,\cite{Arndt1999a,Gerlich2007a,Hackermuller2004a} or at precise
measurements of molecular
properties\,\cite{Berninger2007a,Gerlich2008a}.

Several studies over the last decades  have indicated that the
ionization efficiency is strongly reduced  for large molecules
\,\cite{Schlag1992a,Becker1995a}. In our present experiments we take
up the theme and investigate the formation and ionization of
biomolecular complexes in a mass range up to 7500\,Dalton.

Tagging of large molecules with different chromophores has already
been demonstrated in earlier experiments\,\cite{Hanley2006a,Edirisinghe2006a}. We here explore a similar use of aromatic amino acids and nucleotides in some
different molecular environments: free, embedded in polypeptides and
in large organic clusters.

Organic clusters have attracted increasing interest in their own right.
Charged amino acid clusters such as (Ser$_n$) ~\cite{Myung2004a} or metal complexes such as (Ser$_8$Na)$_{n}$ ~\cite{Takats2003a} have been studied with regard to chiral selectivity in gas phase collisions. Sublimed amino acid clusters ions also led to
gas-phase enantioselective substitution reactions
 between the amino acids~\cite{Yang2006a}.

The interaction with alkali metals may generally result in important configuration changes in DNA and experiments on the interaction between nucleotides and  alkali atoms revealed the preferential coordination of the metals to some nucleotide species~\cite{Koch2002a}. The particular stability of nucleoside quartets as well as the further clustering of these quartets in the presence of alkali atoms has been observed in recent electrospray experiments~\cite{Aggerholm2003}.

Also the radiation resistance of nucleotide clusters~\cite{Schlathoelter2006a}
and the electronic properties of small AgTrp$_n$-clusters~\cite{Compagnon2006a} shifted into the focus of recent research.

In extension and complement to these earlier studies we here explore the integration of alkali and alkaline earth metal atoms into neutral amino acid  clusters and neutral nucleotide clusters as well as the detection of unusually large aggregates in vuv single photon ionization.

\section{Experimental Methods}
\begin{figure}
   \centering
   \includegraphics[width=9cm]{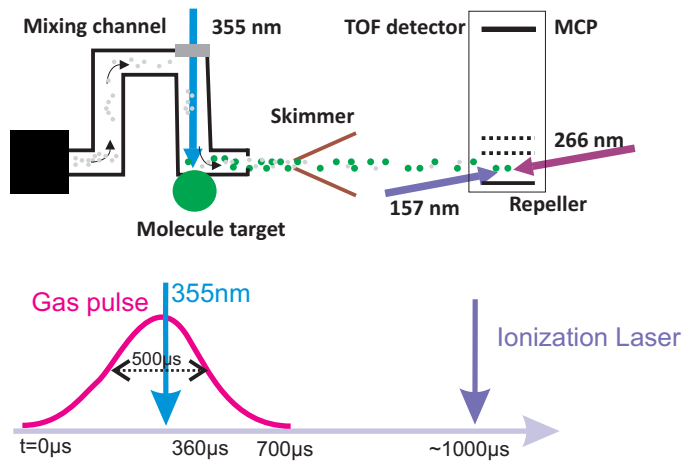}
   \caption{TOP: A mixture of biomolecular powder and cellulose is
   desorbed inside the collision channel. The neutral molecules
   and clusters propagate through a skimmer towards the
   time-of-flight mass spectrometer. A quadrupled Nd:YAG laser
   (266\,nm) and a F$_2$ laser (157\,nm) can be alternatively
   used for the sensitive detection of large organic
   complexes. Bottom: A time line depicts the crucial steps of the experiment: 1) opening of the valve, 2)  desorption shot and 3) ionization pulse.} \label{abb:figure1}
  \end{figure}

The general outline of the experiment is similar to our earlier
studies\,\cite{Marksteiner2008a, Marksteiner2006a} and sketched in
\ref{abb:figure1}: molecular powder is first admixed with cellulose
in a weight ratio of 1:1 to stabilize it against crumbling. It is
then pressed onto a threaded metal rod to form the target, which is
slowly rotated during the desorption process.

The molecules are laser desorbed inside a closed mixing channel
using a pulsed Nd:YAG laser beam ($\mathrm{\lambda=355\,nm,
\tau=5\,ns, E=7-10\,mJ}$). The light beam is focused onto a spot of about
0.5\,mm diameter. A fast valve (0.5\,mm orifice,
500\,$\mu$s opening time) injects a noble gas jet with a typical
backing pressure of 2\,bar into the mixing channel to provide the
required cooling and forward acceleration.
The pulsed beam of neutral molecules passes a
skimmer of 1\,mm diameter before it enters the detection chamber ($10^{-6}$\,mbar). About 50\,cm behind the
source the molecules are photoionized, sorted and detected in a time-of-flight mass spectrometer.

Inside the TOF extraction region, the counter propagating beams of  an $\mathrm{F_2}$ excimer laser ($\tau=5$\,ns, $\lambda$=157\,nm) and of a frequency-quadrupled Nd:YAG laser ($\tau=6-9$\,ns,$\lambda=266$\,nm) cross the molecular beam at right angles.
Both beams were shaped to have circular profiles with a diameter of 5\,mm.
The laser energy is measured directly at the laser exit and we
correct for all losses along the optical path.
The path of the excimer laser is continuously purged with dry nitrogen to ensure optimal transmission of the vuv radiation.
All experiments are performed at a repetition rate of 10\,Hz.

\section{Results}
\subsection{UV vs. VUV photoionization of Trp} We start by comparing
the photoionization mass spectra of tryptophan (Trp) under
irradiation by 266\,nm and 157\,nm laser light. While uv ionization
at 4.66\,eV requires at least a two-photon process, a single vuv
photon at 7.9\,eV can already ionize tryptophan in some
conformational states~\cite{Wilson2006a}.

\begin{figure}
   \centering
   \includegraphics{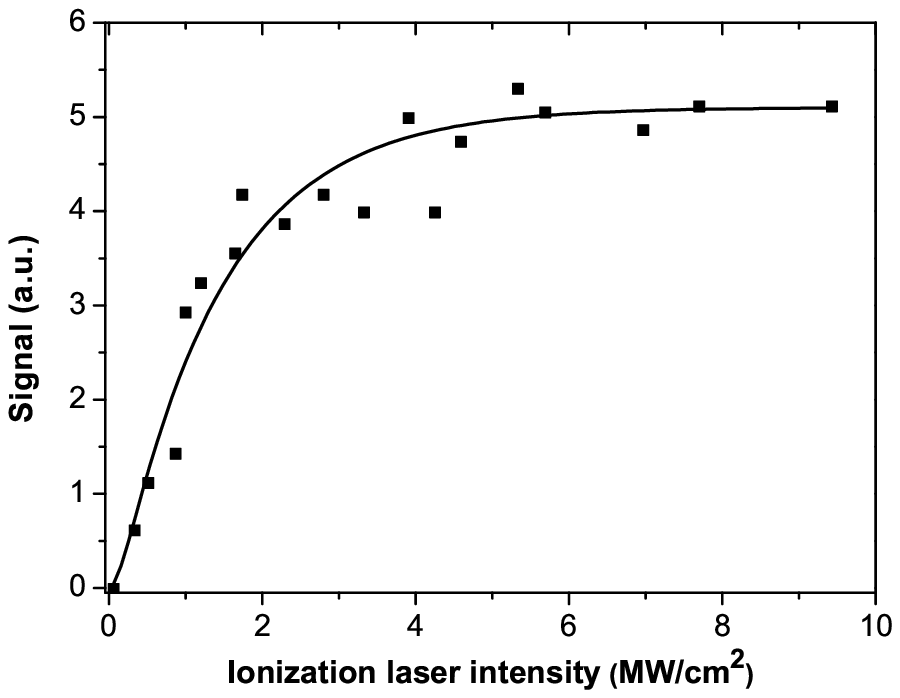}
   \includegraphics{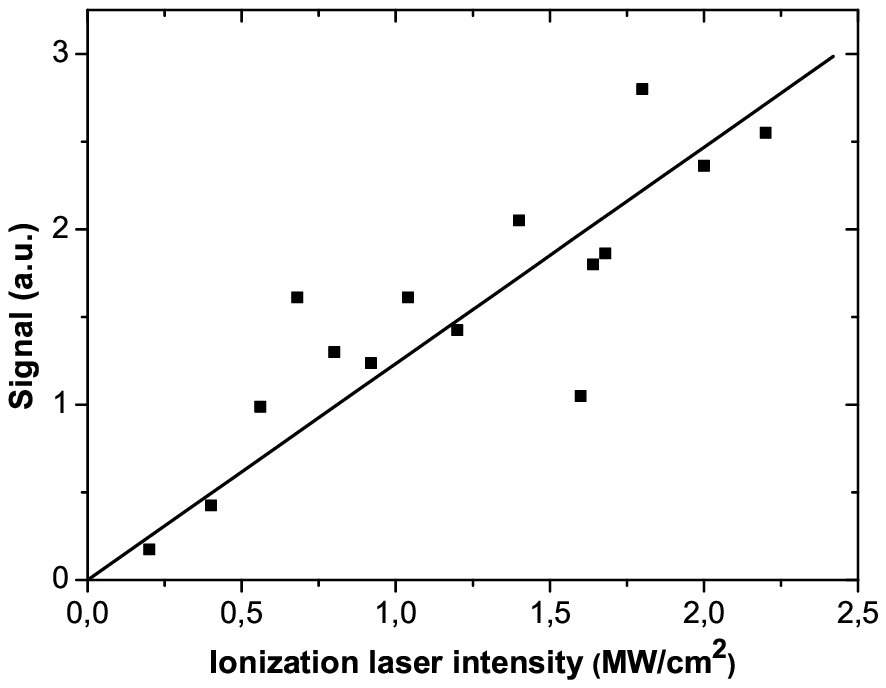}
   \caption{Comparison of the tryptophan ion yield. Top:
   Ionization with uv (266\,nm) light. The saturated curve is fitted by
   Eq.~\ref{equation1}. Bottom: Ionization with vuv
   (157\,nm) light. The straight line is a guide for the eye.}
   \label{abb:figure2}
  \end{figure}

We expect to see this difference also in the intensity dependent ion
yield, which is shown in \ref{abb:figure2}. The 266\,nm signal
increases up to a photon fluence of  $5\times 10^{24}$ s$^{-1}$
cm$^{-2}$, i.e. a laser intensity of
4\,$\mathrm{MW/cm^2}$. The data set
is well represented by a saturation curve for a resonant 2-photon
ionization process~\cite{Witzel1998a}:
\begin{equation}
\begin{split}
S(I)\propto1+\frac{e^{-(s_1+s_2)I\tau/(2h\nu)}(s_1-s_2)}{2s_2}-\\
\frac{e^{-(s_1-s_2)I\tau/(2 h\nu)}(s_1+s_2)}{2s_2},
\end{split}
\label{equation1}
\end{equation}
with $s_1=2\sigma_1+\sigma_2$ and
$s_2=\sqrt{4\sigma_1^2+\sigma_2^2}$. Since tryptophan is ionized in
a resonant two step process, $\sigma_1$ designates the one-photon cross
section for the first transition from the ground state to the intermediate state while
$\sigma_2$ is the coefficient to lift the electron from the intermediate state into the continuum. The laser intensity
is denoted by I, \emph{h} is Planck's constant, $\nu$ is the laser
frequency and $\tau$ is the laser pulse duration.

The observation of
saturation allows us to extract the absorption cross sections $\sigma_1$ and $\sigma_2$, even though with a rather large uncertainty since the two parameters are not fully independent in Eq.~\ref{equation1}~\cite{Witzel1998a}. To fit the tryptophan curve in \ref{abb:figure2}a with
equation\,\ref{equation1} the data were smoothed and an uncertainty
of 20\,\% was taken into account. We thus find the cross sections
 $\sigma_1=1.0(4) \times 10^{-16}\,\mathrm{cm^2}$ and
$\sigma_2=6(4) \times 10^{-16}\,\mathrm{cm^2}$. The systematic uncertainty is
estimated to exceed the statistical value by a factor of two.

The observation of saturation under uv ionization is of practical
importance since we can exploit it to generate a stable ion signal:
Even though the laser energy of flashlamp pumped lasers may exhibit
shot-to-shot fluctuations by 10\% or more, its influence on the
ion counting stability is drastically reduced within the saturation plateau.
For single photon ionization at 157\,nm we expect a linear
dependence of the ion yield as a function of the laser power. \ref{abb:figure2}b shows that this is indeed the case. We find
no saturation within the available laser intensity and here we don't need it since the vuv
excimer laser (Coherent Existar) is internally energy stabilized to
better than 2\%. The ion yield is
comparable at both wavelengths when the laser intensities are kept on equal levels, i.e. when the uv flux is about twice as strong as the vuv flux.
\begin{figure}
   \centering
   \includegraphics{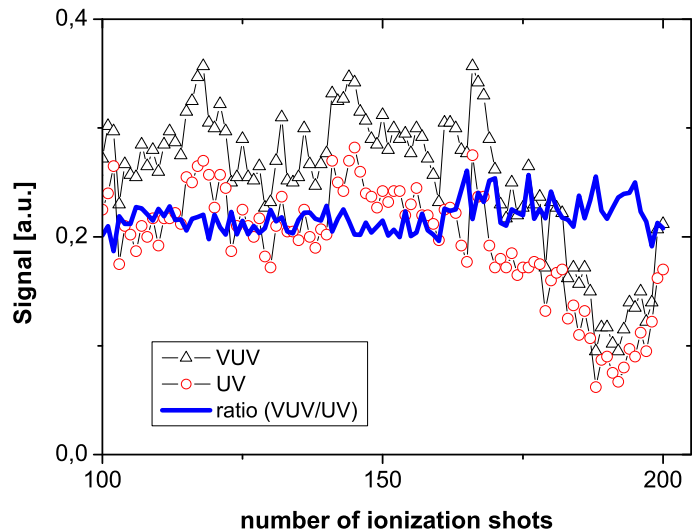}
   \caption{
    The ionization yield shows strong variations, even on a short
    time scale. A normalization of uv (circles) and vuv (triangles) ionization signals,
    recorded with only minimal delay, allows to reduce the
    source-dependent signal fluctuations by one order of
    magnitude. For clarity the normalized signal (uv/vuv = solid
    line) has been scaled by a factor of six.} \label{abb:figure3}
  \end{figure}
On the other hand, laser \emph{desorption} often causes quite
significant fluctuations of the molecular beam intensity. It is not uncommon that signal amplitude variations exceed levels of 50\% within a given laser pulse series, as
shown in \ref{abb:figure3}. This is intolerable in quantitative
experiments and in many cases we rather require a shot-to-shot stability of better than 10\%.

The availability of a second ionization laser pulse can now help us
to reduce the source-related signal fluctuations, which may be caused by
sample inhomogeneities or fluctuations in the desorption
laser intensity. The idea is to derive a normalization signal that varies in
proportion to the desired ion count rate.

For that purpose we use the same molecular desorption pulse twice and send the two ionization laser pulses with a short time delay $\Delta t$ onto the same molecular cloud. The two pulses interact with two neighboring velocity classes in the molecular beam but experiments show that the shape of the velocity distribution is rather well reproducible and that the ratio of
the two subsequent ionization signals is rather constant even if the molecular beam intensity varies from shot to shot.

The time lag  $\Delta t=12\,\mu$s is chosen  to be sufficiently long for the two signals to be separated and stored by the same TOF-MS pulse. It is
also sufficiently short for both lasers to interact with nearly the
same velocity class ($\Delta v/v < 2\%$).

The normalization procedure can be further refined by using
different laser beam diameters to interact with different parts of
the molecular beam. In our normalization experiments the vuv laser beam
was set to 5\,mm whereas the uv beam was vertically focused to below
1\,mm.

In~\ref{abb:figure3} we show the raw data of the tryptophan ion signal
after vuv and uv excitation. The signals were
recorded in two box car integrators. The ratio of both
curves (vuv/uv) is plotted as the continuous line of~\ref{abb:figure3}. While the statistical error of the raw data may
even vary by more than 50\% the normalized signal is steadily stabilized to
better than 6\%. Normalization can thus efficiently eliminate source-related
fluctuations by about one order of magnitude.

\subsection{VUV ionization of polypeptides and organic clusters}
The successful ionization of tryptophan with vuv light indicates that this soft detection scheme could be useful also for larger organic molecules and
clusters with an ionization energy below 8\,eV. Theoretical studies along these lines have been performed by Dehareng et al.~\cite{Dehareng2004a} and experiments with Phe-Gly-Gly and beta-carotene have been presented by Wilson et al.~\cite{Wilson2006a}. We
here extend these studies to larger polypeptides as well as to a variety of nucleotide clusters and metal-amino acid complexes.

\subsubsection{Gramicidin and biomolecular clusters} Gramicidin
ionizes well and softly, both under uv and vuv irradiation. In both
cases the mass spectrum shows predominantly the pure parent peak
centered at 1884\,Dalton, and a small forest of peaks below 300
Dalton, but no fragments in between.

\begin{figure}
   \centering
   \includegraphics{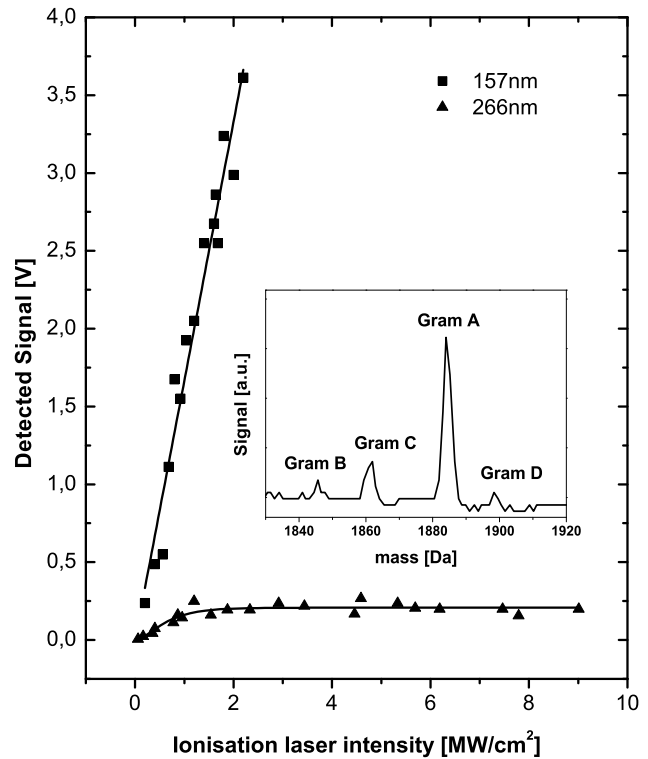}
   \caption{Ion yield of gramicidin A (1884\,Da) as a
   function of the laser intensity at 266\,nm and 157\,nm, respectively. The
   157\,nm curve is well-represented by a straight line, whereas the 266\,nm curve
   follows a saturated 2-photon ionization curve
   (Eq.~\ref{equation1}). The inset shows a gramicidin spectrum recorded after uv ionization. All variants from Gra A to Gra D are visible.} \label{abb:figure4}
  \end{figure}

Commercial gramicidin powder consists of a mixture of the four
polypeptides gramicidin A, B, C and D. Their amino acid sequence is
HCO-\textbf{X}-Gly-L-Ala-D-Leu-L-Ala-D-Val-L-Val-D-Val-L-Trp-D-Leu-L-\textbf{Y}-D-Leu-L-Trp-D-Leu-L-Trp-NHCH$_2$CH$_2$OH.
Gramicidin A, B and C differ only by a single amino acid at residue
11 (Y), which can be either tryptophan\,(A), phenylalanine\,(B) or
tyrosine\,(C). Gramicidin\,D is derived from gramicidin\,A by
substituting alanine in place of glycine as the first residue (X).
In our powder the composition was nominally A=80-85\%, B=6-7\%,
C=5-14\%, D$<1$\,\% (Sigma Aldrich). All four gramicidin molecules
can be detected with both ionization wavelengths (see~\ref{abb:figure4} inset).

Although the mass spectra are rather similar for both excitation
wavelengths, the ion yields can differ dramatically, as shown in
\ref{abb:figure4}. The vuv ion signal of intact gramicidin A
(1884\,Da) outperforms the uv ionization already by a factor of fifteen (!),
at the available vuv laser intensity of
2.2\,$\mathrm{MW/cm^2}$. This increase is surprisingly high. The uv absorption spectra of tryptophan and gramicidin in solution\,\cite{Gotsche2007a} are rather similar and suggest that gramicidin actually ionizes via the tryptophan residues that it contains. This argument would rather lead us to expect the ionization yields of gramicidin and tryptophan to have rather the same spectral characteristics. On the other hand, the single-photon ionization efficiency of tryptophan strongly depends on the molecular conformation and only a small subset of all its possible conformations has a vertical ionization energy that can be accessed by the F$_{2}$-laser~\cite{Dehareng2004a}. Our observation is therefore consistent with the hypothesis that the integration of the tryptophan residues into the polypeptide chain actually favor stabilization of tryptophan conformations with a lower ionization potential.

At 157\,nm the ion yield of gramicidin shows an unsaturated linear increase with the laser intensity. In contrast to that, the
266\,nm ionization curve runs again into saturation, as also
observed for tryptophan. To fit the saturated curve with
equation\,\ref{equation1} the recorded data were treated as in the case of tryptophan before and a fit with equation\,\ref{equation1}
results in $\sigma_1=6(2) \times 10^{-16}\,\mathrm{cm^2}$ and
$\sigma_2=3(1) \times 10^{-16}\,\mathrm{cm^2}$. As expected, the gramicidin cross sections are larger than those for tryptophan but the systematic uncertainties are comparable.

\begin{figure}
   \centering
   \includegraphics{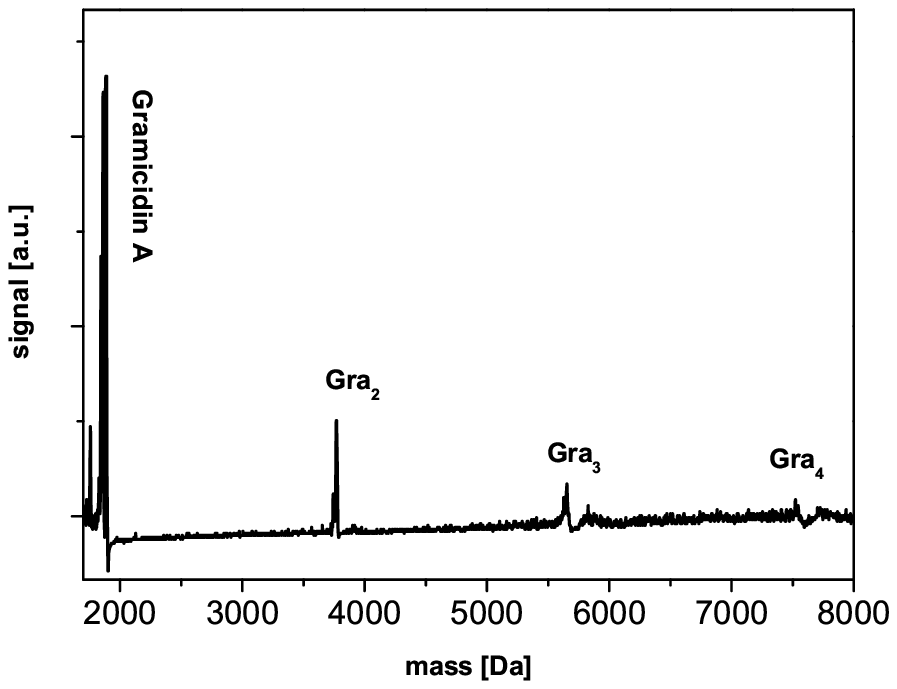}
   \includegraphics{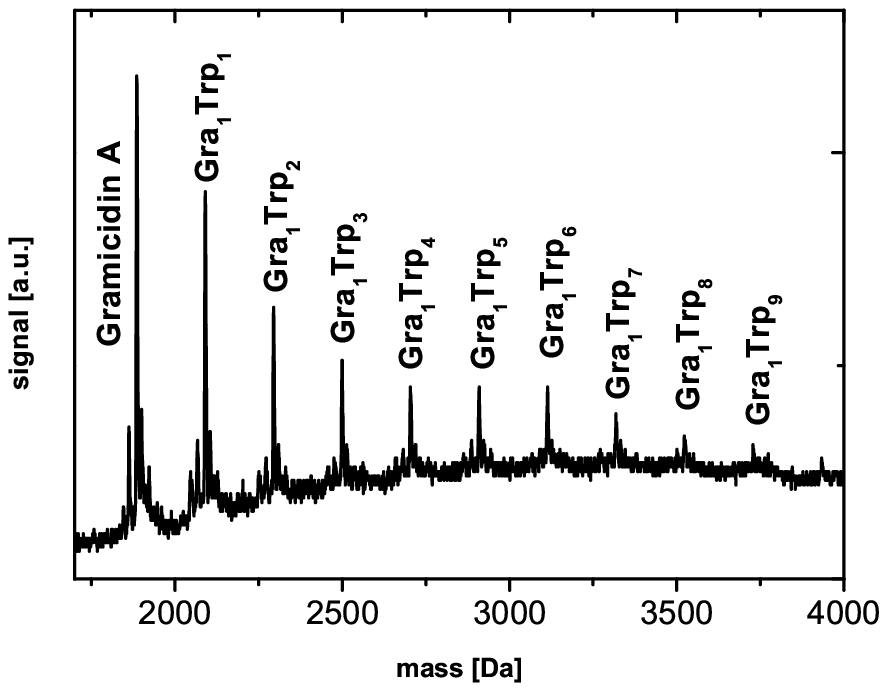}
   \includegraphics{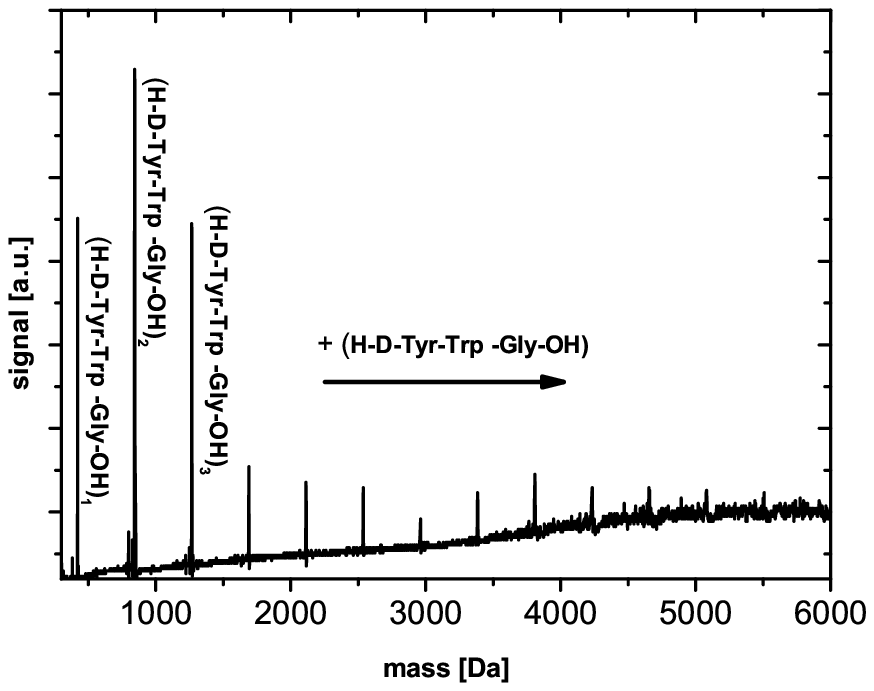}
   \caption{VUV ionization of massive polypeptide clusters. Top:
   Neutral gramicidin cluster. Center:  tryptophan-gramicidin clusters with admixed
   insulin. Note: although insulin was
   admixed to the same sample, we detect neither any pure
   insulin nor any trpytophan-tagged insulin under vuv ionization. Bottom: Desorption of H-D-Tyr-Trp-Gly-OH (425\,Da)
   leads to large neutral clusters which can still be detected up
   to a size of
    (H-D-Tyr-Trp-Gly-OH)$_{13}$ at a mass of 5500\,Dalton.}
   \label{abb:figure5}
  \end{figure}

It turns out that vuv ionization is so efficient that also clusters
of gramicidin (Gra$_n$) and mixed amino acid-polypeptide clusters,
such as Gra$_1$Trp$_n$ can be observed. \ref{abb:figure5}\, (top)
shows the mass spectrum after the desorption of a gramicidin-cellulose mixture (1:1). Interestingly, only pure gramicidin clusters form. The gramicidin
dimer\,(3768\,Da), trimer(5652\,Da) and even the
neutral tetramer at 7536\,Da can still be detected.

During the experiments with pure Gramicidin clusters a linear mixing channel with 4\,mm diameter exit hole was used. For all other runs we used the meandering desorption tube as shown in the introduction and in ~\ref{abb:figure1}.

But also when using the meandering channel we find a number of mixed clusters
when we co-desorb gramicidin with tryptophan, insulin and cellulose
in a mixture of 0.8:1:1.2:1.6. The mass spectrum is shown in
\ref{abb:figure5}\,(center). While cluster formation comes not
unexpectedly, it is worth noting that \ref{abb:figure5}\,(middle)
shows only the signal of the gramicidin monomer followed by a number
of peaks which correspond to the sequential addition of single
tryptophan molecules. Up to nine tryptophan molecules could be
attached to one single gramicidin molecule.

The successful binding between the polypeptide gramicidin and
tryptophan as a tag with a low ionization threshold raised
hopes that such as dye-tagging might also be fruitful for the
detection of neutral larger biomolecules, such as insulin. The
spectra showed, however, no indication whatsoever of insulin,
insulin-tryptophan clusters or derivatives thereof.

Complementary experiments with the tripeptide H-D-Tyr-Trp-Gly-OH
(425\,Da, Bachem Inc.), confirmed that also these particles form large
clusters. The tripeptide was desorbed from a highly diluted mixture
with the metal salt $\mathrm{CaCO_3}$ and cellulose (1:40:100). And
yet, postionization with vuv light revealed a cluster
distribution leading up to (H-D-Tyr-Trp-Gly-OH)$_{13}$, as shown in
\ref{abb:figure5}\,(bottom). In spite of the enormous relative
abundance of calcium and cellulose in the sample, the tripeptide
clusters remain always pure. This is in marked difference to
tryptophan complexes which readily embed a single calcium atom
~\cite{Marksteiner2008a}.

The abundance distribution of the tripeptide is non-monotonic but we
can post-ionize and detect clusters up to about 6000\,Da. This is
comparable in size to insulin which remained undetected in all our
experiments.

\begin{figure}
   \centering
   \includegraphics{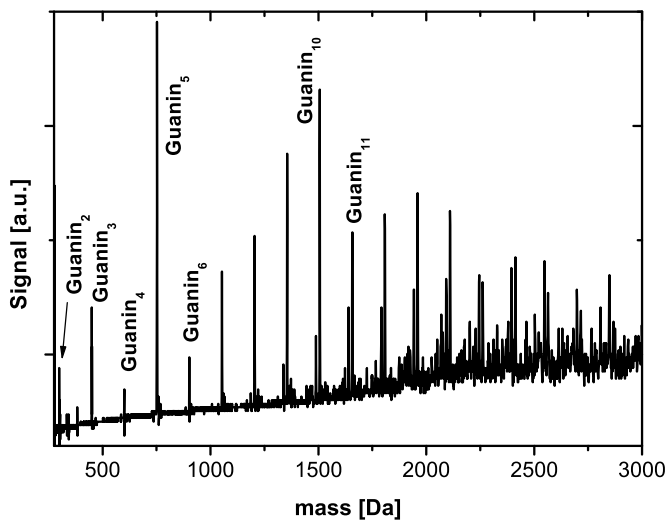}
   \caption{Guanine clusters form during laser desorption into
   an argon seed gas and ionized by vuv light. The pentamer G$_{5}$ clearly dominates the spectrum.
   The absence of Ca-complexes is conspicuous, in particular in comparison
   to the amino acid experiments reported below, where the integration of a single Calcium atom is strongly favored.} \label{abb:figure6}
  \end{figure}

One may ask how the clustering characteristics change if we replace
the amino acids by nucleotides. This question has been addressed
using a mixture of guanine (151\,Da), CaCO$_3$ and cellulose
(mixture: 1:1:0.7). The result is shown in \ref{abb:figure6}. The
total intensity of all guanine clusters varies from shot to shot,
but G$_5$ always stands out as particularly abundant, 'magic', cluster.
It is still an open question whether this phenomenon is related to the particular stability of
G$_5$ or to a favorable tuning of the ionization energies in this particular cluster.
The influence of photo-induced fragmentation processes seems to be rather small, as the
photon energy matches already the ionization threshold\,\cite{Zavilopulo2009a} and there is hardly any excess energy deposited in the clusters.

The observation of the magic G$_5$ cannot be explained by fragmention during the ionization process, since the ionization energy of Guanine amounts to the vuv photon energy.
Magic cluster size were also found in electrospray ionization experiments with guanine and other nucleotides, however starting from ions in electrospray experiments~\cite{Koch2002a}.
It is furthermore interesting to note that we never observe the
inclusion of either calcium or carbonates or cellulose fragments in
this series. In case of the guanine we also repeated the
experiment without any metal salts in the sample. We find the same cluster distribution as before. In spite of their abundance, the metal atoms obviously play no role in the formation of the guanine clusters. The role of the metal salts in catalyzing large biomolecular complexes is also discussed in the following section.

\subsubsection{Amino acid metal complexes} 
\begin{figure}
   \centering
   \includegraphics{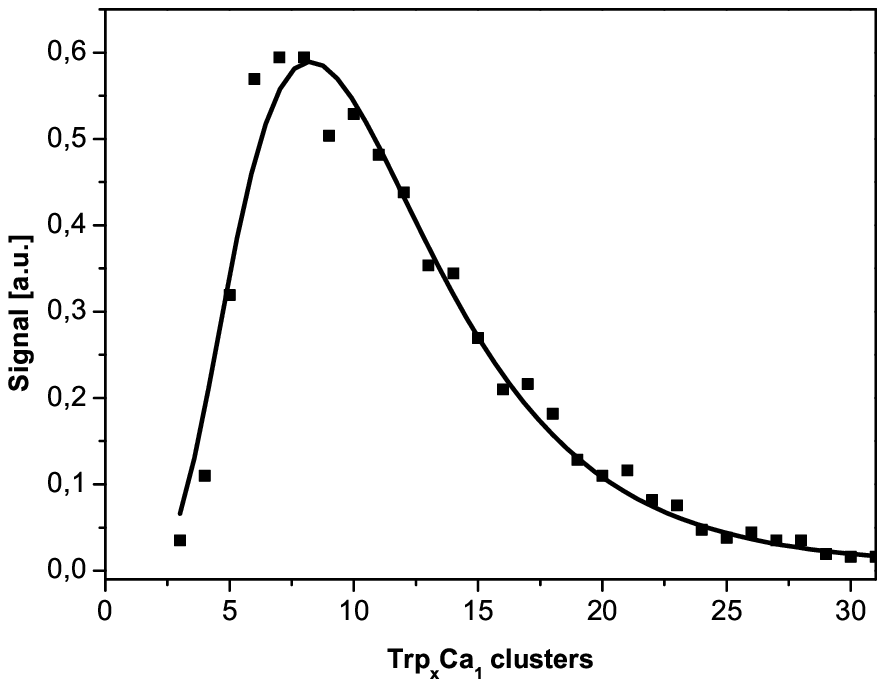}
   \includegraphics{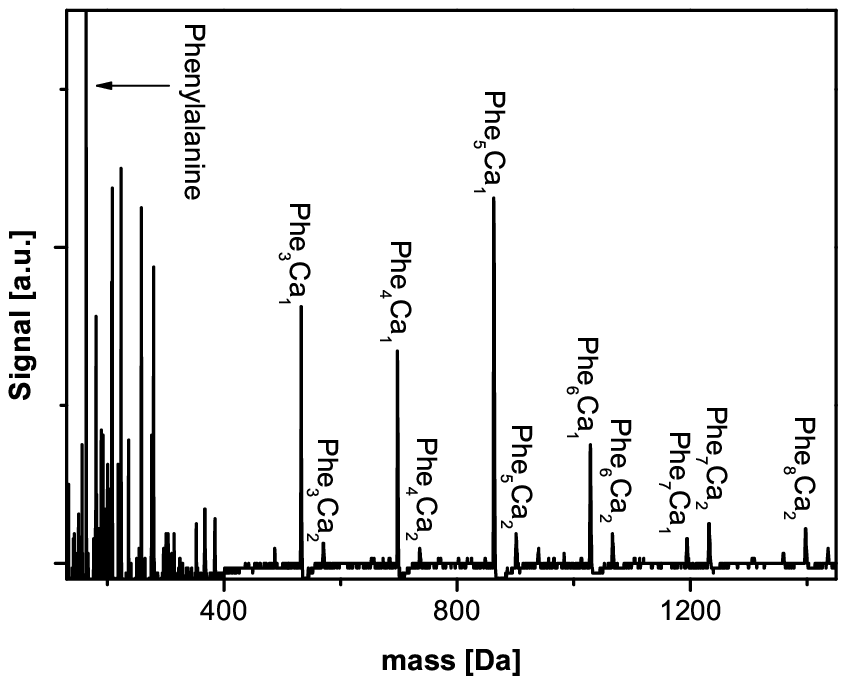}
   \caption{Neutral metal-organic complexes form after laser
   desorption from a mixture of amino acid powder with a metal salt
   and cellulose. Top: Cluster distribution of Trp$_x$Ca$_1$ resulting from desorbed
   tryptophan and CaCO$_3$ fitted with a log-normal distribution.
   Ionization of the clusters was done with vuv light and argon was used seed gas.
   Bottom: Phenylalanine and CaCO$_3$ ionized with uv radiation
   (backing gas: argon)~\cite{Marksteiner2008a}.} \label{abb:figure7}
  \end{figure}

The detection of
gramicidin and the tripeptide clusters is rather efficient but
photoionization of single large molecules beyond 2000~Da seems
generally to be a great challenge. It has been shown in the previous
section that the formation and postionization of biomolecular clusters
is possible up to a mass of 7500\,Da (see \ref{abb:figure5} top).
Also the presence of alkali earth salts can assist in the formation
of large tryptophan clusters, such as Trp$_n$Ca$_1$ complexes
\cite{Marksteiner2008a}, which could still be photo-detected beyond a
mass of 6000\,Dalton (\ref{abb:figure7} top).

It is natural to ask if metal catalyzed cluster formation also
applies to other amino acids, such as for instance Phenylalanine
(Phe) which shares its aromatic character with tryptophan. The
analogy between the two molecules does, however, not extend so far
as to allow detection using the same laser frequency. VUV excitation
does not result in any measurable Phe ion signal neither for the
monomer nor for any higher cluster.

The ionization energy of Phenylalanine, was determined to be 8.6\,eV
~\cite{Wilson2006a} and exceeds the energy of a single photon at
157\,nm. A vuv two-photon process is not impossible but it would
deposit extra internal energy and fragmentation would
become more likely~\cite{Plekan2008a}. Accordingly, isolated
molecules with an ionization potential exceeding 8\,eV, such as
phenylalanine or histidine, were never observed in our setup under
vuv excitation -- whereas they readily appeared after 266\,nm
two-photon ionization.

\ref{abb:figure7}\,(bottom) shows the mass spectrum of Phenylalanine
that was first desorbed from a balanced mixture of Phe, CaCO$_3$ and
cellulose  (ratio 1:1:1) and then ionized at 266\,nm. We see the
single Phe molecule at 165\,Da, as well as Phe complexes that
contain preferentially one and less frequently two calcium atoms. In
contrast to the nucleotide guanine which does not incorporate any
metal atom at all, even in the over-abundant presence of CaCO$_3$,
the amino acid Phenylalanine prefers to incorporate a singe metal
atom, independent of its actual cluster size. This is in agreement
with earlier studies of Trp$_n$Ca$_1$
complexes~\cite{Marksteiner2008a}.
\begin{figure}
   \centering
   \includegraphics{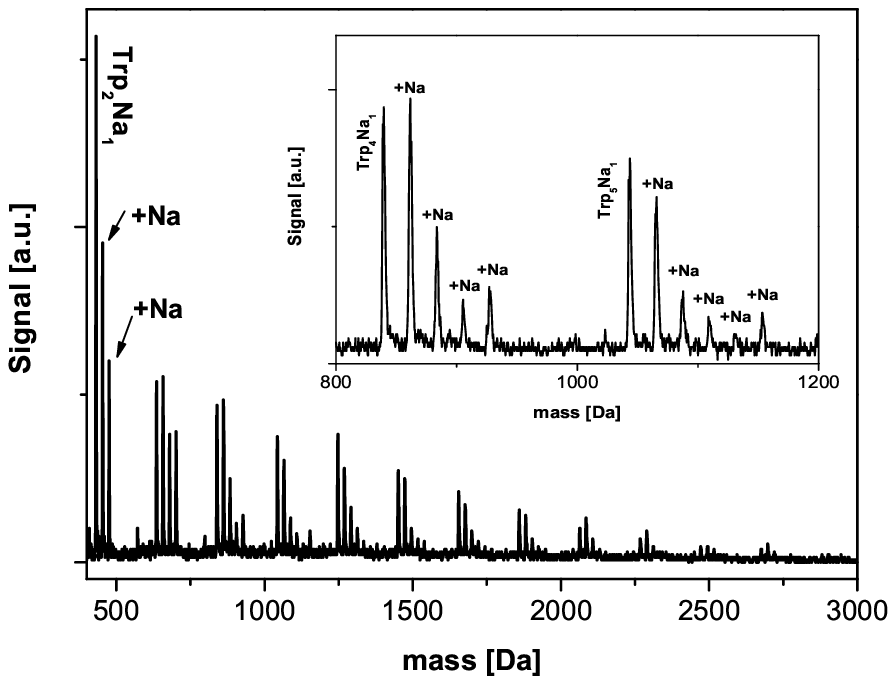}
   \includegraphics{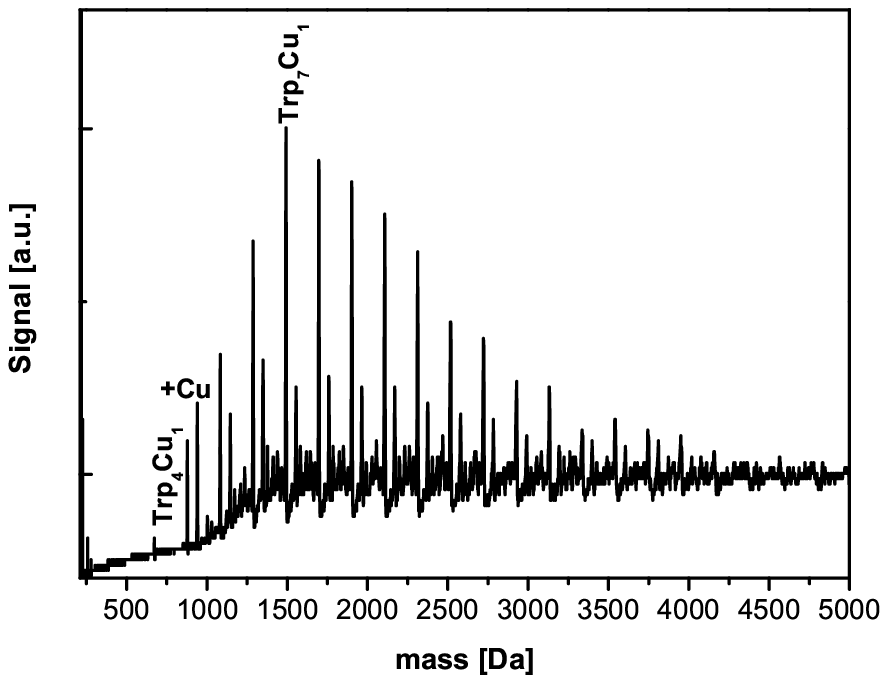}
   \caption{Neutral metal-organic complexes forming after laser
   desorption from a mixture of the biomolecular powder, the
   metal salt and cellulose. The complexes were ionized using VUV light. a) tryptophan and Na$_2$CO$_3$
   (backing gas:Neon) b) tryptophan and CuCO$_3$ (backing gas:
   Argon) } \label{abb:figure9}
  \end{figure}
We finally pose the question how the substitution of alkaline earth
salts by alkali or copper salts will influence the amino acid
complex formation. \ref{abb:figure9}\,(top) shows the mass spectrum
after co-desorption of tryptophan, Na$_2$CO$_3$ and cellulose
(mixture: 1:0.3:1.6). We observe dominantly complexes of the form
Trp$_n$Na$_{1}$ ... Trp$_n$Na$_{n+1}$ as shown in
\ref{abb:figure9}\,(top). This rule is valid at least up to
Trp$_5$Na$_6$ and only violated by the absence of Trp$_5$Na$_5$. In
the given mass range, the signals are clearly not yet limited by the
signal-to-noise ratio, nor by the availability of metal atoms in
the plume. We therefore conclude that the maximum number of metal
atoms per cluster is rather determined by the nature of the bonds.
A better understanding of the observed cluster rules requires future numerical
modeling.

Replacing the metal salt by CuCO$_3$ also leads to the formation of
tryptophan-copper clusters, as can be seen in
\ref{abb:figure9}\,(bottom). The cluster formation starts with
Trp$_4$Cu$_1$ and every Trp$_x$Cu$_n$ cluster can contain one or two
copper atoms (n=1,2). With the exception of
Trp$_4$Cu$_1$ the signal for Trp$_n$Cu$_1$ is always stronger than
for  Trp$_n$Cu$_2$.

Our experiments showed that the production and detection of
metal-biomolecular complexes can be very efficient. In case of
tryptophan this led again to complexes up to 6000\,Da which were
generated and/or detected more readily than clusters without metal
inclusions.

When assessing the cluster spectra one should, however, also bear in
mind, that we never see the neutral clusters directly but only after
postionization in a single-photon process. The incorporation of
metal atoms can generally enhance the optical absorption cross
section, as has recently been shown for small tryptophan-silver
complexes\,\cite{Compagnon2006a}.

The abundance distribution of metal-biomolecule clusters is
therefore also influenced by modifications of the absorption cross
sections and the ionization cross sections.

The metal atoms may also contribute to the cluster stability or the
cluster aggregation time. Generally, they might do this
catalytically - i.e. without being finally included -- or as part of
the resulting complex.

\section{Conclusions}
The cluster physics or metal-organic system is a rich field. It
can be accessed using laser desorption of neutral samples,
buffer gas assisted aggregation and uv/vuv
photoionization.

In our experiments we study the formation and detection of
neutral massive clusters and metal-organic complexes which are built from
amino acids, nucleotides and polypeptides in combination with alkali
and alkali-earth metals.

We find in particular that vuv ionization can be very efficient and
ionizes the polypeptide gramicidin up to fifteen times more
efficiently than radiation at 266\,nm. Gramicidin can thus be
observed even up to the neutral tetramer at 7536\,Da.

Desorption of a mixture of tryptophan, gramicidin and insulin shows
that gramicidin-tryptophan clusters can be generated and seen even
up to Gra$_1$Trp$_9$. In contrast to that, dye-tagging of neutral
insulin with tryptophan led to no detectable signals:
Insulin-Tryptophan complexes were never recorded in our machine.
This is in marked difference to tripeptide (H-D-Tyr-Trp-Gly-OH)
nucleotide clusters which formed easily detectable aggregates up to
masses well comparable to that of insulin.

Our experiments show that metal atom-amino acid complexes can be
readily produced, when metal-carbonates are co-desorbed with organic
molecules. Calcium, sodium and copper atoms were shown to be readily
embedded into tryptophan complexes but they seem to be consistently
excluded even by large tripeptide complexes (built from aggregations of H-D-Tyr-Trp-Gly-OH) or by Guanin clusters.

The cluster abundance distribution shows a rich structure in almost
all of the spectra. Further work is still required to understand the
observation of 'magic' cluster numbers, and to attribute the
relative importance of metal inclusions for either cluster
stabilization or cluster detection, as well as to elucidate the kinetics of
formation, the temperature dependence and the stability of these large
complexes. In this context it is particularly interesting to study
the conspicuous differences in cluster catalysis for different
biomolecules in the presence of different metal salts.

Our experiments thus raise many questions which will require
further experiments and a thorough theoretical treatment
of large complexes. The answers to some of the many open questions
may eventually also come from matter wave interferometry, which was
the starting point of our own efforts in this field. Recent
experiments have shown that near-field interferometry has the
potential to provide us with precise static~\cite{Berninger2007a} or
optical polarizabilities~\cite{Gerlich2008a} or absolute molecular
spectra~\cite{Nimmrichter2008b} - which in turn may serve as
important benchmark values for future molecular cluster models.

\section*{Acknowledgments}
The authors thank Christoph Dellago and Harald Oberhofer for
fruitful discussions. This work has been supported by the Austrian
Science Funds FWF within the Wittgenstein program  Z149-N16.

\providecommand*{\mcitethebibliography}{\thebibliography}
\csname @ifundefined\endcsname{endmcitethebibliography}
{\let\endmcitethebibliography\endthebibliography}{}

\end{document}